\documentclass[aps,pra,showpacs,twocolumn,superscriptaddress,a4paper]{revtex4}

\usepackage{ulem}
\usepackage[usenames]{color}

\usepackage{graphics}

\begin{document}


\title{Effects of periodic potentials on the
critical velocity \\ of superfluid Fermi gases in the BCS-BEC crossover
}



\author{Gentaro Watanabe}
\affiliation{Asia Pacific Center for Theoretical Physics (APCTP), POSTECH,
San 31, Hyoja-dong, Nam-gu, Pohang, Gyeongbuk 790-784, Korea}
\affiliation{Nishina Center, RIKEN, 2-1 Hirosawa, Wako, Saitama 351-0198, Japan}
\author{Franco Dalfovo}
\affiliation{INO-CNR BEC Center and Department of Physics, University of Trento, 38123 Povo, Italy}
\author{Lev P. Pitaevskii}
\affiliation{INO-CNR BEC Center and Department of Physics, University of Trento, 38123 Povo, Italy}
\affiliation{Kapitza Institute for Physical Problems, 119334 Moscow, Russia}
\author{Sandro Stringari}
\affiliation{INO-CNR BEC Center and Department of Physics, University of Trento, 38123 Povo, Italy}

\date{\today}

\begin{abstract}
We study the effects of an external periodic potential on the critical
velocity of a superfluid Fermi gas in the crossover between
the Bardeen-Cooper-Schrieffer (BCS) phase and Bose-Einstein
condensation (BEC). We numerically solve the Bogoliubov-de Gennes
equations to model a three-dimensional (3D) gas of ultracold atoms in
the superfluid phase flowing through a 1D optical lattice. We find
that when the recoil energy is comparable to the Fermi energy, the
presence of the periodic potential reduces the effect of pair-breaking
excitations. This behavior is a consequence of the peculiar band 
structure of the quasiparticle energy spectrum in the lattice.  When 
the lattice height is much larger than the Fermi energy, the periodic 
potential makes pairs of atoms to be strongly bound even in the BCS 
regime and pair-breaking excitations are further suppressed.
We have also found that when the recoil energy is comparable to
or larger than the Fermi energy, the critical velocity due to 
long-wavelength phonon excitations shows a non-monotonic 
behavior along the BCS-BEC crossover.

\end{abstract}

\pacs{03.75.Ss, 03.75.Lm, 67.85.De}

\maketitle

\section{Introduction}

Ultracold atom gases in optical lattices have been continuously 
attracting great interest for the last ten years 
\cite{morsch,lewenstein,bloch}. Recent developments in the field of 
ultracold atom gases provide a new research arena in the physics of 
quantum fluids: by using Feshbach resonances of ultracold Fermi atoms, 
one can study the crossover from the Bardeen-Cooper-Schrieffer (BCS) phase 
to a Bose-Einstein condensate (BEC) of molecules \cite{giorgini}.
In the current research frontier, superfluidity of ultracold Fermi gases
in optical lattices is very intriguing problem, which has
interesting connections with similar issues in solid state physics, 
nuclear physics, and astrophysics.

The critical velocity of superflow due to energetic instability
is one of the most important properties of superfluids, which has been
pioneered by Landau \cite{landau}. If the velocity of superflow
exceeds some critical value, the kinetic energy of the superfluid 
can be dissipated by creating excitations 
\cite{landau,nozieres_pines,pethick_smith,pitaevskii_stringari}.
In uniform superfluid Fermi gases in the BCS-BEC crossover,
excitations which cause the energetic instability are 
of two types: fermionic pair-breaking excitations in the BCS regime 
and  long-wavelength phonon excitations in the BEC regime
\cite{andrenacci,combescot}. In the unitary regime both 
mechanisms are suppressed and the critical velocity shows a maximum 
value \cite{andrenacci,combescot,sensarma}.

Recently, effects of periodic potentials on the critical velocity
of Fermi superfluids has been studied experimentally \cite{miller}.
This experiment has stimulated theoretical investigations 
of this problem \cite{burkov,vc,ganesh,yunomae}. Most of them 
has focused on the BCS regime in tight-binding approximation 
\cite{burkov,ganesh,yunomae}. The purpose of the present work is 
to obtain an understanding of the critical velocity from a unified 
point of view covering all regions along the BCS-BEC crossover and 
both the strong and weak lattice regime. To this purpose, we use the 
Bogoliubov-de Gennes (BdG) equations. This theory accounts for both  
types of excitations which are relevant in this problem. In our 
previous work \cite{vc} we already used it for a gas at unitarity; 
here we extend the calculations in order to explore the whole 
crossover region. As a main result, we find that, when the lattice 
height is comparable to or much larger than the Fermi energy, the 
periodic potential reduces the effect of pair-breaking
excitations.  This is due to the periodic structure 
of the quasiparticle energy spectrum in the Brillouin zone and the 
formation of the bound molecules induced by the lattice. 
Another main result is that when the recoil energy is comparable to
or larger than the Fermi energy, the critical velocity due to 
long-wavelength phonon excitations shows a non-monotonic 
behavior along the BCS-BEC crossover.
These effects are unique for Fermi superfluids in periodic potentials 
and do not exist in the case of single barrier potentials 
\cite{camerino,vc,camerino_review}.

This paper is organized as follows.  In Sec. \ref{formalism},
we explain the basic formalism employed in the present work.
Then we show the results in Sec.\ \ref{result}.
Finally, summary and outlook are given in Sec.\ \ref{summary}.

\section{Basic Formalism\label{formalism}}

We want to study the effect of the periodic potential on 
the Landau critical velocity of Fermi superfluids in the 
whole BCS-BEC crossover, in situations where the Fermi 
energy is larger or smaller than the lattice height. 
For this aim, we need to use a theoretical framework which 
can account for the formation of bound molecules induced by the 
periodic potential, which is important when the lattice height is 
larger than the Fermi energy \cite{orso,optlatunit}; the same 
formalism must also account for pair-tunneling processes, which 
are important on the BEC side of the resonance \cite{optlatunit,ohashi}.  
A suitable approach consists of the numerical solution of 
the Bogoliubov-de Gennes (BdG) equations
\cite{bdg}:
\begin{equation}
\left( \begin{array}{cc}
H'(\mathbf r) & \Delta (\mathbf r) \\
\Delta^\ast(\mathbf r) & -H'(\mathbf r) \end{array} \right)
\left( \begin{array}{c} u_i( \mathbf r) \\ v_i(\mathbf r)
\end{array} \right)
=\epsilon_i\left( \begin{array}{c} u_i(\mathbf r) \\
v_i(\mathbf r) \end{array} \right) \; ,
\label{eq:BdGnonuniform}
\end{equation}
where $u_i$ and $v_i$ are quasiparticle amplitudes and
$\epsilon_i$ the corresponding eigen-energies. The single-particle
hamiltonian is  $H'(\mathbf r) =-\hbar^2 \nabla^2/2m +V_{\rm ext}-\mu$,
where $m$ is the atom mass and $V_{\rm ext}(\mathbf r)$ is the 
external potential.  The order parameter (or gap parameter) 
$\Delta (\mathbf r)$ and 
the chemical potential $\mu$, appearing in Eq.\ (\ref{eq:BdGnonuniform}), 
are variational parameters determined from the gap equation,
\begin{equation}
  \Delta(\mathbf r) =-g \sum_i u_i(\mathbf r) v_i^*(\mathbf r) \; ,
\label{eq:gap}
\end{equation}
together with the constraint 
\begin{equation}
  n_0=\frac{2}{V}\sum_i \int \left| v_i(\mathbf r) \right|^2d{\bf r} \; ,
\end{equation}
enforcing the conservation of the average density $n_0$. 
Here $g$ is the coupling constant for the contact interaction and 
$V$ is the volume of the system. The BdG eigenfunctions 
obey the normalization condition $\int d^3r \left[u_i^*({\bf r})u_j({\bf
r}) +v_i^*({\bf r})v_j({\bf r})\right]=\delta_{i,j}$.
Finally, the energy density $e$ can be calculated as
\begin{equation}
e = \frac{1}{V}\int \! d{\bf r} \sum_i [ 2(\mu-\epsilon_i )|v_i({\bf r})|^2
+ \Delta^*({\bf r}) u_i({\bf r})v_i^*({\bf r})]  .
\label{eq:energydens}
\end{equation}

In the present study, we consider a three-dimensional superfluid
Fermi gas, which is uniform in the $x$ and $y$ directions and subject to
a one-dimensional optical lattice along $z$:
\begin{equation}
  V_{\rm ext}(z)=sE_{\rm R} \sin^2q_{\rm B}z \equiv V_0 \sin^2q_{\rm B}z \; .
\label{lattice}
\end{equation}
Here $V_0\equiv sE_{\rm R}$ is the lattice height,
$s$ is the laser intensity in dimensionless units, $E_{\rm
R}=\hbar^2q_{\rm B}^2/2m$ is the recoil energy, $q_{\rm B}=\pi/d$ is
the Bragg wave vector, and $d$ is the lattice constant.
For practical reasons, throughout this paper, we set $s=1$ 
except for special cases, which we shall mention explicitly. The
ratio between the Fermi energy and the lattice height 
is then varied by changing the average density of the gas.

In the presence of a supercurrent with wave vector $Q=P/\hbar$
moving in the direction of the periodic potential, 
one can write the gap parameter in the form
\begin{equation}
  \Delta(\mathbf r)=e^{i 2Q z}\tilde{\Delta}(z) ,
\label{eq:bloch}
\end{equation}
where $\tilde{\Delta}(z)$ is a complex function with period $d$. 
Therefore, from the gap equation, we see that the eigenfunctions of 
Eq.~(\ref{eq:BdGnonuniform}) must have the Bloch form $u_i(\mathbf r) =
\tilde{u}_i(z) e^{i Q z}e^{i\mathbf k \cdot \mathbf r }$ and
$v_i(\mathbf r) = \tilde{v}_i(z) e^{-i Q z}e^{i\mathbf k \cdot
\mathbf r }$. The wave vector $k_z$ lies in the first 
Brillouin zone (i.e., $-q_{\rm B} \le k_z \le q_{\rm B}$)
and $\tilde{u}_i$ and $\tilde{v}_i$ are periodic in 
$z$ with period $d$. We also define the quasi-momentum 
$P_{\rm edge}$ and quasi-wavenumber $Q_{\rm edge}$ at the edge of 
the Brillouin zone for $P$ and $Q$ as $P_{\rm edge}=\hbar Q_{\rm edge}\equiv 
\hbar q_{\rm B}/2$. 
(Note that the edge of the first Brillouin zone
for $P$ is at $P_{\rm edge} =q_{\rm B}/2$ while that for $k_z$ is 
at $q_{\rm B}$.)
This Bloch decomposition transforms 
Eq.~(\ref{eq:BdGnonuniform}) into the following BdG equations 
for $\tilde{u}_i$ and $\tilde{v}_i$:
\begin{equation}
\left( \begin{array}{cc}
\tilde{H}'_{Q}(z) & \tilde{\Delta}(z) \\
\tilde{\Delta}^\ast(z) & -\tilde{H}'_{-Q}(z) \end{array} \right)
\left( \begin{array}{c} \tilde{u}_i(z) \\ \tilde{v}_i(z)
\end{array} \right)
=\epsilon_i\left( \begin{array}{c} \tilde{u}_i(z) \\
\tilde{v}_i(z) \end{array} \right) \;,
\label{eq:BdGnonuniform2}
\end{equation}
where
\begin{equation}
  \tilde{H}'_{Q}(z)\equiv \frac{\hbar^2}{2m} \left[k^2_x+k^2_y
+\left(-i\partial_z+Q+k_z\right)^2\right]+V_{\rm ext}(z) -\mu\, .
\label{hq}
\end{equation}
Here, the label $i$ represents the wave vector $\mathbf k$ as well 
as the band index. In order to remove the ultraviolet 
divergences in the BdG equations with contact potentials, we use 
the regularization scheme proposed by Refs.~\cite{bruun,bulgac}.
Since we need to calculate the second derivatives of the 
energy with respect to the density and the quasi-momentum,
we use large values of the cutoff energy $E_C$,
especially in the BEC side, where the size of the pair is 
much smaller than the average inter-atomic distance 
(details are described in Sec.~\ref{result}).

As discussed in Refs.~\cite{andrenacci,combescot}, 
the energetic instability of superfluids of dilute Fermi gases 
can be caused by two processes \cite{note_assumption}: 
the creation of long-wavelength superfluid phonon 
excitations or fermionic pair-breaking excitations.
The critical velocity by the former process can be determined by
the hydrodynamic analysis of the excitations 
\cite{machholm,taylor,pitaevskii,pethick_smith,vc}.  
Starting from the continuity equation and the Euler equation, and
linearizing with respect to the perturbations of the density and 
the velocity fields, we obtain the dispersion relation of the 
long-wavelength phonon,
\begin{equation}
\omega (q) =\frac{\partial^2 e}{\partial n_0 \partial P}\ q
+\sqrt{\frac{\partial^2 e}{\partial n_0^2}
\frac{\partial^2 e}{\partial P^2}}\ |q|\ .
\label{eq:dispersion}
\end{equation}
Here, $\hbar\omega$ and $q$ are the energy and the wavenumber of 
the excitations,
$n_0$ and $P$ are the average density and the quasimomentum
of the superfluids.
The energetic instability occurs when $\omega(q)$ becomes negative:
\begin{equation}
\frac{\partial^2 e}{\partial n_0 \partial P} =
\sqrt{\frac{\partial^2 e}{\partial n_0^2}
\frac{\partial^2 e}{\partial P^2}}\ .
\label{eq:vcrit-hydro}
\end{equation}
In practice, we calculate the energy density 
$e(n_0,P)$ for a given $n_0$ and $P$ from Eq.\ (\ref{eq:energydens})
using the solution of the BdG equations (\ref{eq:BdGnonuniform2}).
Then the critical quasimomentum $P_c$ at which the energetic instability occurs
is determined by Eq.\ (\ref{eq:vcrit-hydro}) \cite{note_partial}.
We finally obtain the critical velocity $v_c$ from
\begin{equation}
  v_c = \frac{1}{n_0}\left(\frac{\partial e}{\partial P}\right)_{P_c} .
\label{eq:vc}
\end{equation}

\begin{figure}[tbp]
\begin{center}\vspace{0.0cm}
\rotatebox{0}{
\resizebox{8.5cm}{!}
{\includegraphics{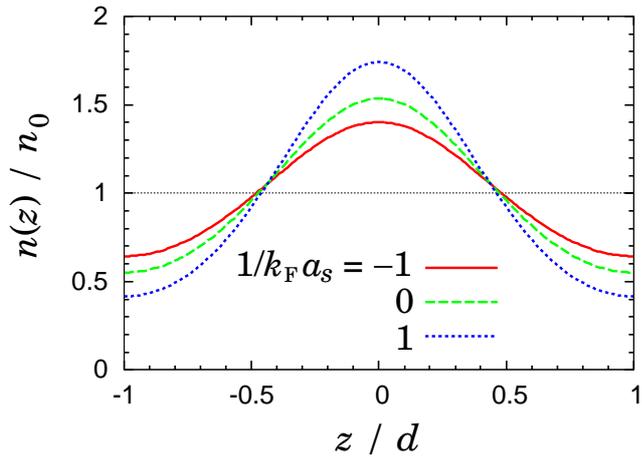}}}
\caption{\label{fig:n}(Color online)\quad
Density profile $n(z)$ at $P=0$ in the BCS-BEC crossover 
for $E_{\rm F}/E_{\rm R}=1$ and $s=1$.
Here, $n(z)$ is normalized by the average density $n_0$
}
\end{center}
\end{figure}

On the other hand, the critical velocity due to the
pair-breaking fermionic excitations can be determined by
looking at the quasiparticle energy spectrum 
$\epsilon_i$. The energetic instability by the pair-breaking 
excitations occurs when some quasiparticle energy $\epsilon_i$ 
starts to be negative: 
\begin{equation}
\epsilon_i\le 0\ .
\label{eq:pb}
\end{equation}
From Eq.~(\ref{eq:vc}) evaluated at the critical quasi-momentum 
determined by this condition, we obtain a critical velocity
for the pair-breaking excitations. The actual critical velocity of
the system is the lowest between the ones obtained from the above
two conditions \cite{Lev}. We finally note that the gas becomes 
unstable also when some excitation energy starts to have a 
non-zero imaginary part. This corresponds to a dynamical 
instability, which causes an exponential growth of the
amplitude of the perturbation. To address 
the problem of dynamical instability, short-wavelength bosonic 
excitations should be also properly included. This is beyond the 
scopes of the present work, in which we instead focus on the 
energetic instability. Results of the critical velocity 
for dynamical instability due to long-wavelength excitations
are given in Appendix \ref{dynamical}.

\section{Results\label{result}}

We study the three cases of 
$E_{\rm F}/E_{\rm R}=2.5$, $1$, and $0.1$ 
with a fixed value of $s=1$ except for a few cases which we 
shall mention explicitly. 
Here $E_{\rm F} = \hbar^2k_{\rm F}^2/(2m)$ and $k_{\rm F}=(3\pi^2 n_0)^{1/3}$
are the Fermi energy and momentum, respectively, of a uniform
noninteracting Fermi gas of density $n_0$.
For each value of $E_{\rm F}/E_{\rm R}$, we solve the BdG 
equations for several values of the parameter $1/k_{\rm F}a_s$
along the crossover from the BCS to the BEC side, namely  
$1/k_{\rm F}a_s=-1$, $-0.5$, $0$, $0.5$, and $1$, where 
$a_s$ is the $s$-wave scattering length of atoms.

In the $x$ and $y$ directions, we assume periodic boundary conditions
with period $L_\perp = 12\pi k_{\rm F}^{-1}$.
We set the cutoff energy $E_C$ as follows:
for $E_{\rm F}/E_{\rm R}=2.5$, $E_C=40 E_{\rm F}$ in the BCS side
and $E_C=100 E_{\rm F}$ at unitarity and in the BEC side;
for $E_{\rm F}/E_{\rm R}=1$, $E_C=50 E_{\rm F}$ in the BCS side
and $E_C=100 E_{\rm F}$ at unitarity and in the BEC side;
for $E_{\rm F}/E_{\rm R}=0.1$, $E_C=250 E_{\rm F}$ in the BCS side,
$E_C=350 E_{\rm F}$ at unitarity, and $E_C=500 E_{\rm F}$ in the BEC side.

\begin{figure}[tbp]
\begin{center}
\rotatebox{0}{
\resizebox{8.2cm}{!}
{\includegraphics{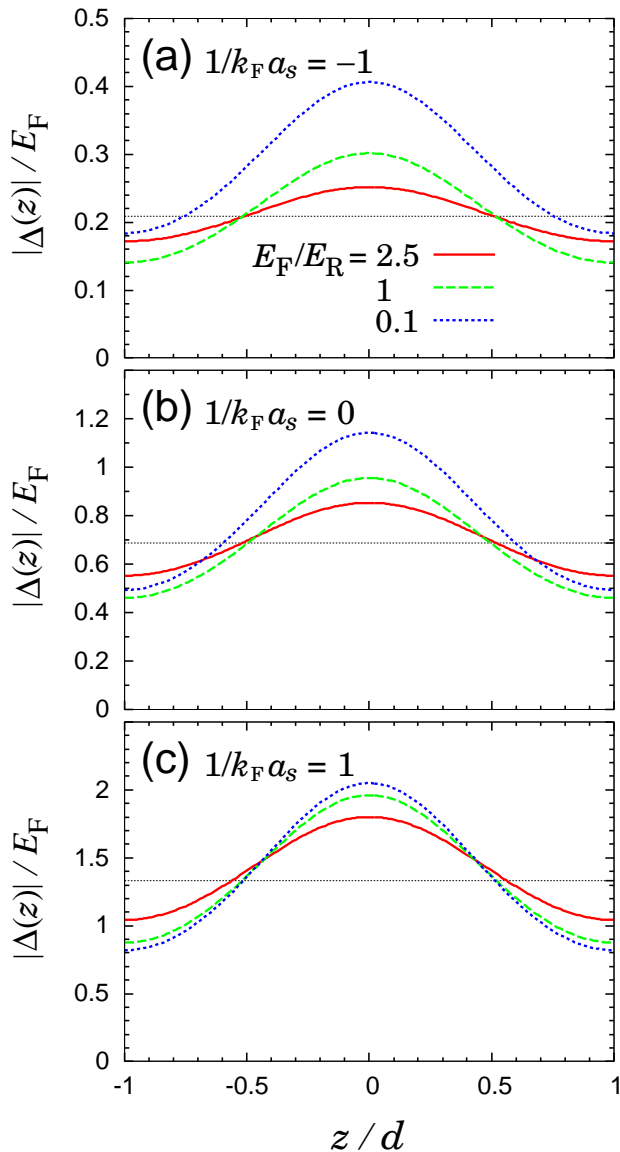}}}
\caption{\label{fig:delta}(Color online)\quad
Amplitude $|\Delta(z)|$ of the gap parameter
at $P=0$ in the BCS-BEC crossover: 
$1/k_{\rm F}a_s=-1$ (a), 0 (b), and 1 (c).
The horizontal dotted lines show the amplitude of the gap parameter
for the uniform system at the same value of $1/k_{\rm F}a_s$.
}
\end{center}
\end{figure}

\subsection{Density profiles and gap parameter}

\begin{figure*}[tbp]
\begin{center}\vspace{0.0cm}
\rotatebox{270}{
\resizebox{9.9cm}{!}
{\includegraphics{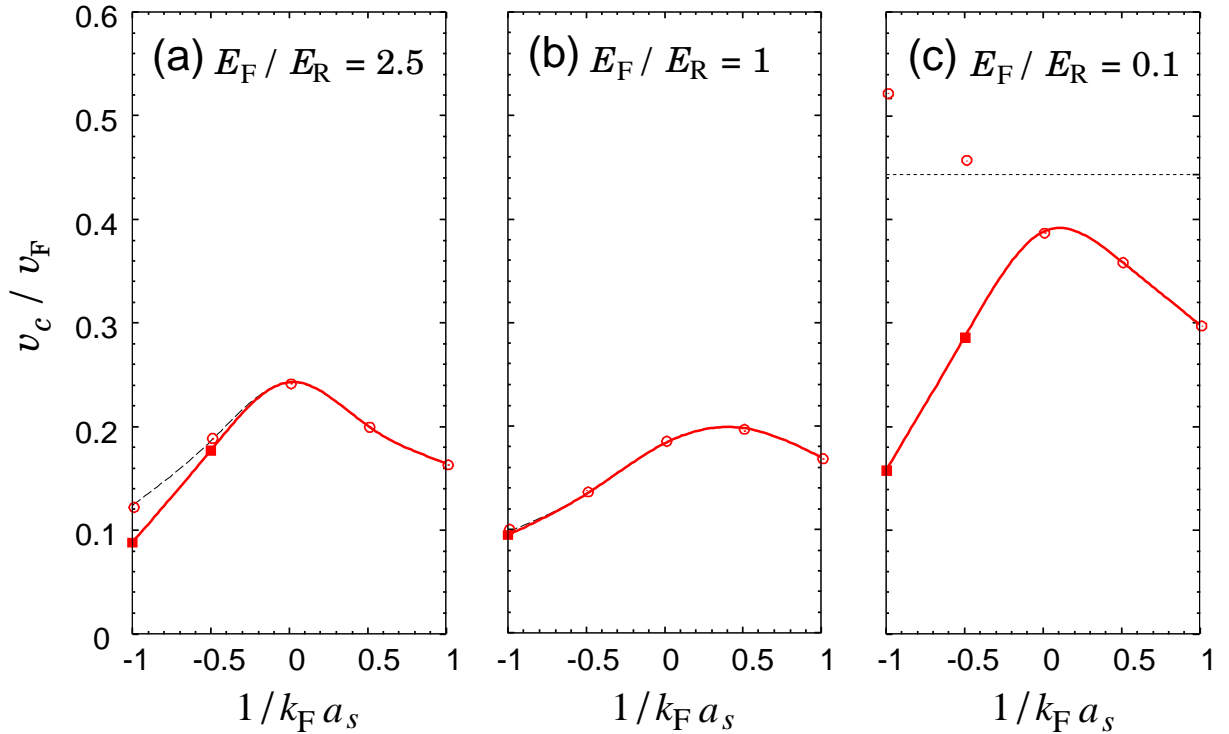}}}
\caption{\label{fig:vc}(Color online)\quad
Critical velocity $v_c$ of the energetic instability
for $E_{\rm F}/E_{\rm R}=2.5$ (a), $1$ (b), and $0.1$ (c) 
with $s=1$ in the BCS-BEC crossover.
Open circles and filled squares show the critical velocity 
due to long-wavelength phonons and fermionic pair-breaking excitations,
respectively.
The horizontal dotted line in panel (c) represents the 
value of the sound velocity
$c_s^{(0)}$ of uniform system at unitarity, 
$c_s^{(0)}/v_{\rm F}=(1+\beta)^{1/2}/\sqrt{3}\simeq 0.443$.
The red solid lines and the black dashed lines
are guides to the eye.
}
\end{center}
\end{figure*}

\begin{figure}[tbp]
\begin{center}\vspace{0.0cm}
\rotatebox{0}{
\resizebox{8.5cm}{!}
{\includegraphics{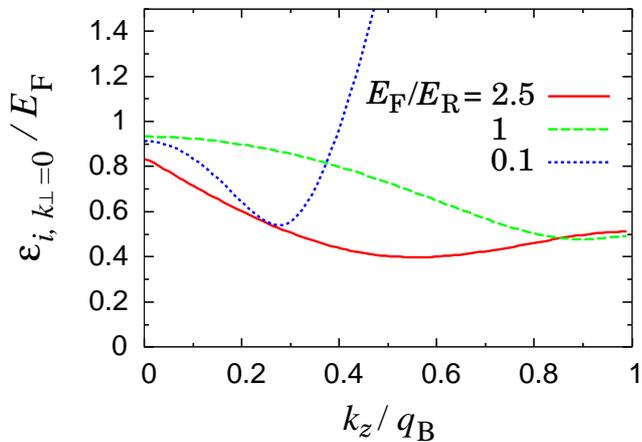}}}
\caption{\label{fig:qp_energy}(Color online)\quad
Lowest band of the quasiparticle energy spectrum $\epsilon_i$ 
for $P=0$ and $1/k_Fa=-0.5$.
Here, we show the first radial branch with 
$k_\perp^2 \equiv k_x^2+k_y^2=0$.
For $E_{\rm F}/E_{\rm R}=1$, the minimum of $\epsilon_i$ is located 
close to the Brillouin zone edge $k_z=q_{\rm B}$.
}
\end{center}
\end{figure}

\begin{figure}[tbp]
\begin{center}\vspace{0.0cm}
\rotatebox{0}{
\resizebox{8.2cm}{!}
{\includegraphics{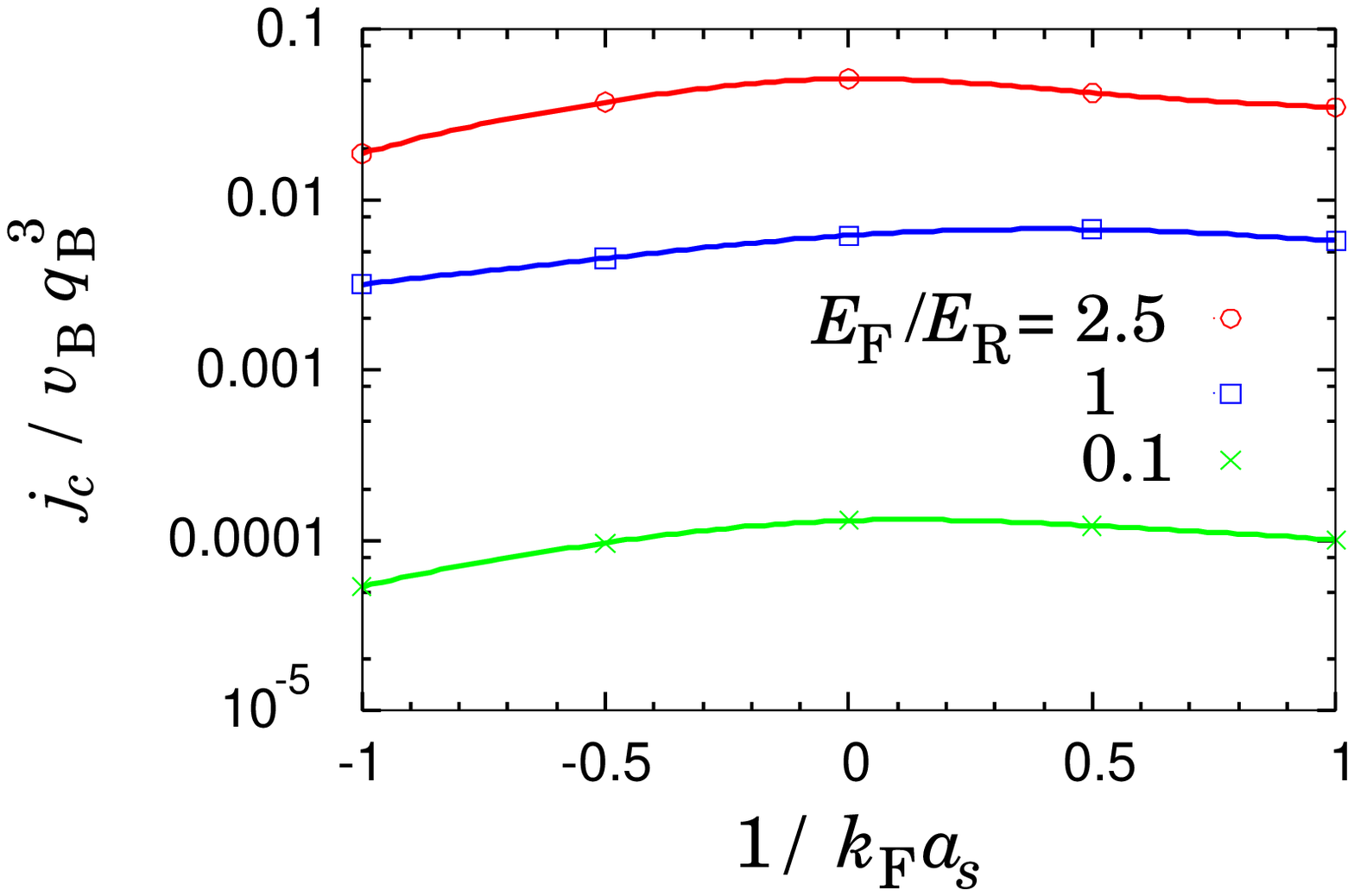}}}
\caption{\label{fig:jc}(Color online)\quad
Critical current $j_c=n_0 v_c$ for the same cases 
in Fig.~\protect\ref{fig:vc}.  Here we show the lowest value of $j_c$,
given either by long-wavelength phonon excitations
or by fermionic pair-breaking excitations.
The curves connecting symbols are guides to the eye.}
\end{center}
\end{figure}

In Fig.~\ref{fig:n}, we show the density profile
$n(z)=2\sum_i \left| \tilde{v}_i(z) \right|^2$ of Fermi atoms at $P=0$ 
along the BCS-BEC crossover for $E_{\rm F}/E_{\rm R}=1$. Moving from the 
BCS regime ($1/k_{\rm F}a_s=-1$) to the BEC 
regime ($1/k_{\rm F}a_s=1$), the density $n(z)$ becomes more 
inhomogeneous. This behavior is consistent with the fact that  
the compressibility of a uniform Fermi gas is known to 
increase monotonically with $1/k_{\rm F}a_s$. We observe the same 
qualitative behavior for the other values of $E_{\rm F}/E_{\rm R}$.

In Fig.~\ref{fig:delta}, we show the amplitude $|\Delta(z)|$ of 
the order parameter at $P=0$ for different values 
of $1/k_{\rm F}a_s$ and $E_{\rm F}/E_{\rm R}$. From this figure 
one can see that, especially in the BCS regime ($1/k_{\rm F}a_s=-1$),
the order parameter $|\Delta|$ is enhanced when the
Fermi energy is smaller than the lattice strength, as shown by
the blue curves for $E_{\rm F}/E_{\rm R}=0.1$ 
(i.e., $E_{\rm F}/V_0=0.1$). We understand this fact 
as due to the formation of bosonic molecules induced by 
the external periodic potential.  This process is indeed 
expected to become significant when the lattice is strong 
\cite{optlatunit}.

\subsection{Critical velocity}

Before presenting the numerical results for the critical 
velocity, let us discuss the conditions upon which the periodic 
potential can produce significant effects on the behavior of 
long-wavelength phonons and pair-breaking excitations. 
For long-wavelength phonons the condition is 
$E_{\rm F}/E_{\rm R}\agt 1$ or $s\gg 1$. In fact, in the opposite
case, $E_{\rm F}/E_{\rm R}\ll 1$ and $s\alt 1$, the right-hand 
side of Eq.~(\ref{eq:vcrit-hydro}), which coincides
with the sound speed at $P=0$, is much smaller than the 
maximum value of the left-hand side of Eq.~(\ref{eq:vcrit-hydro}), 
which is of order $q_{\rm B}/m$ (see also Fig.~5 in Ref.~\cite{vc}).
Thus Eq.~(\ref{eq:vcrit-hydro}) is satisfied at $P\simeq 0$
and the critical velocity is basically determined by the sound 
speed at $P=0$, which is also close to the sound speed in the 
uniform system. Consequently, the critical velocity due to 
long-wavelength phonons is almost unaffected by the presence of 
the lattice if $E_{\rm F}/E_{\rm R}\ll 1$ and $s\alt 1$,  
even though the lattice height $V_0$ is large compared to the 
Fermi energy $E_{\rm F}$. On the other hand, the sufficient 
condition for pair-breaking excitations to be affected by the 
periodic potential due to the formation of bound molecules is 
$E_{\rm F}/V_0\ll 1$. This condition can be satisfied either 
by decreasing $E_{\rm F}/E_{\rm R}$ or by increasing $s$.  
When $E_{\rm F}/E_{\rm R}\simeq 1$,  the critical velocity for 
pair-breaking excitations is also affected by a peculiar band 
structure of the quasi-particle spectrum.

Our results for the critical velocity $v_c$ are shown in 
Fig.~\ref{fig:vc} for $s=1$. Let us first concentrate 
on the results at high density, $E_{\rm F}/E_{\rm R}=2.5$, in 
panel (a). The open circles correspond to the critical velocity
for long-wavelength phonons, which exhibits a non-monotonic behavior.
In particular, in the BCS regime (negative $1/k_{\rm F}a_s$) this 
critical velocity is strongly reduced compared to the one in a
uniform gas (see, e.g., Fig.~8 in Ref.~\cite{combescot}). This 
is a peculiar effect of the lattice. However, for the parameters
of Fig.~\ref{fig:vc}(a), the actual critical velocity in the BCS 
regime is still given by fermionic pair-breaking excitations (filled 
squares). The latter are almost unaffected by the lattice and 
therefore, the overall behavior of the critical velocity in the 
crossover is qualitatively similar to that of a uniform gas, 
already discussed in Ref.~\cite{combescot}: in the BCS regime, 
$v_c$ increases when approaching unitarity ($1/k_{\rm F}a_s =0$), 
because the intra-pair attraction becomes stronger and thus the 
amplitude of the gap parameter increases; in the opposite BEC 
regime ($1/k_{\rm F}a_s>0$), the critical velocity is given 
by long-wavelength phonons and an increase of the inter-pair 
repulsion leads to a larger sound speed and, again, a critical 
velocity $v_c$ increases towards unitarity. 
As a consequence, $v_c$ takes a maximum value at 
$1/k_{\rm F}a_s \simeq 0$.

When the recoil energy is comparable to the Fermi energy,
the periodic potential causes qualitative changes in the 
results of the critical velocity. For $E_{\rm F}/E_{\rm R}=1$ 
[Fig.~\ref{fig:vc}(b)], we observe that, at $1/k_{\rm F}a_s=-0.5$,
the critical velocity is given by long-wavelength phonon excitations
rather than pair-breaking excitations even in the BCS regime.
We understand this effect as mainly due to a peculiar band structure 
of the quasiparticle energy spectrum. In Fig.~\ref{fig:qp_energy}, 
we show the lowest band of the quasiparticle energy spectrum 
$\epsilon_i$ for the first radial branch with 
$k_\perp^2\equiv k_x^2+k_y^2=0$ at $P=0$ and $1/k_Fa=-0.5$.
In general, the quasiparticle spectrum near the center of the 
Brillouin zone, at $|k_z|\simeq 0$, is only weakly affected 
by the periodicity of the system and hence the change of 
$\epsilon_i$ with increasing $P$ is close to that in the uniform 
system, given by the Doppler term $P\hbar k_z/m$. 
On the other hand, close to the zone edge,
at $k_z\simeq \pm q_{\rm B}$, the change of 
$\epsilon_i$ with increasing $P$ is much smaller than 
$P\hbar k_z/m$ because of the periodicity of the Brillouin zone
($\epsilon_i$ at $k_z=\pm q_{\rm B}$ must be identical).
In the case of $E_{\rm F}/E_{\rm R}=1$, the minimum of $\epsilon_i$ 
is indeed located close to the edge of the Brillouin zone, 
unlike the other two cases of $E_{\rm F}/E_{\rm R}=2.5$ and $0.1$. 
Therefore, the reduction of the minimum
value of $\epsilon_i$ with increasing $P$ is relatively small
for $E_{\rm F}/E_{\rm R}=1$ and this is why we find that
$v_c$ is determined by phononic instead of fermionic
excitations in this case.

For smaller density ($E_{\rm F}/E_{\rm R}=0.1$), we observe a
significant increase of $v_c$ in the whole crossover 
[Fig.~\ref{fig:vc}(c)]. 
In our previous article \cite{vc}, we 
already showed that, in a unitary Fermi superfluid with 
$E_{\rm F}/E_{\rm R}\ll 1$, the phononic critical velocity is
almost unaffected by the lattice, if $s\alt 1$, and remains 
close to the speed of sound of a uniform gas with the 
same density \cite{note_sound}. This can be understood 
by recalling that phonons always have wavelength larger than the 
healing length of the superfluid, which is the order of 
$k_{\rm F}^{-1}$ or greater. When $E_{\rm F}/E_{\rm R}\ll 1$,
the healing length becomes much larger than the lattice spacing 
$d =\pi/q_{\rm B}$, and this makes the phonons insensitive to 
the lattice itself.  In the present work, we find that the same 
is true even away from unitarity, leading to a larger critical 
velocity in the whole crossover [empty circles in 
Fig.~\ref{fig:vc}(c)]. The critical velocity due to  
pair-breaking excitations (filled squares) is also increased 
because a lattice strength $V_0$ much larger than the Fermi energy
gives a stronger attraction between paired atoms.

In Fig.~\ref{fig:jc} we show 
the critical current $j_c=n_0 v_c$  for the same cases of 
Fig.~\ref{fig:vc}. Due to the low density at 
$E_{\rm F}/E_{\rm R}=0.1$, the critical current is much 
smaller than the other cases even though $v_c$ in units 
of $v_{\rm B}=q_{\rm B}/m$ for $E_{\rm F}/E_{\rm R}=0.1$
is comparable to that of $E_{\rm F}/E_{\rm R}=1$.

\begin{figure}[tbp]
\begin{center}\vspace{0.0cm}
\rotatebox{0}{
\resizebox{8.5cm}{!}
{\includegraphics{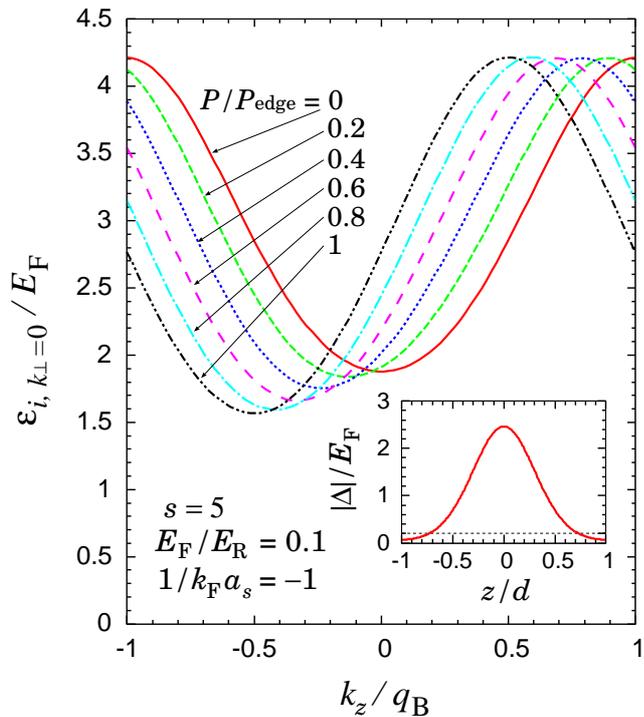}}}
\caption{\label{fig:qp_energy2}(Color online)\quad
Lowest band of the quasiparticle energy spectrum $\epsilon_i$ 
for large lattice height with $s=5$ and $E_{\rm F}/E_{\rm R}=0.1$
(i.e., $E_{\rm F}/V_0=0.02$)
in the BCS regime at $1/k_{\rm F}a_s=-1$.
Here, we show the first radial branch with 
$k_\perp^2 \equiv k_x^2+k_y^2=0$, which always gives the smallest values
of $\epsilon_i$ in this case.
The inset shows the amplitude $|\Delta(z)|$ of the order parameter
at $P=0$. The horizontal dotted line shows the amplitude of the order 
parameter for the uniform system at the same values of $1/k_{\rm F}a_s=-1$.
}
\end{center}
\end{figure}

\subsection{Dependence on the lattice height}

All results shown in Figs.~\ref{fig:n}-\ref{fig:jc} have been
obtained by fixing the lattice height $s=1$ and vary the density in
order to change the key parameter $E_{\rm F}/E_{\rm R}$. 
If we increase $s$ keeping the average density fixed, 
the superfluid flow is suppressed and $v_c$ is also reduced in general.

A systematic analysis as a function of $s$, which would be natural 
from the experimental viewpoint, is computationally very demanding 
and is beyond the scopes of this work. The choice of $s=1$ is not 
accidental, however. It turns out, in fact, that around $s=1$ the 
effects of the lattice on the critical velocity are the most 
pronounced as far as the interplay between pair-breaking and 
long-wavelength bosonic excitations is concerned. For lower 
values of $s$ these two types of excitations behave qualitatively 
the same as in a uniform superfluid, as function of $1/k_Fa$, 
being scarcely affected by the lattice, at least within the range 
of $E_{\rm F}/E_{\rm R}$ considered in this work. On the other hand, 
at larger $s$ the pair-breaking instability is quickly suppressed 
and long-wavelength excitations becomes dominant along the
crossover. 

The reason why a strong lattice prevents the pair-breaking processes
can be understood by looking at Fig.~\ref{fig:qp_energy2}, where we 
show the quasiparticle energy spectra at various values of $P$ in 
the case of $s=5$, $E_{\rm F}/E_{\rm R}=0.1$ (i.e., $E_{\rm F}/V_0=0.02$), 
and $1/k_{\rm F}a_s=-1$.  The spectrum for $P=0$ shows a quadratic 
dependence of $k_z$ with a positive curvature around $k_z=0$ and 
there are no minima at $k_z\ne 0$. Even though the figure represents 
a case in the deep BCS regime, the structure of $\epsilon_i$ is 
consistent with the formation of bound pairs. In the inset of the 
same figure, we show the amplitude $|\Delta(z)|$ of the gap parameter 
at $P=0$. First, we note that the minimum value of 
$|\Delta(z)|$ at $z/d=\pm 1$ is smaller than, but still comparable 
to the value of $|\Delta|$ in the uniform case, suggesting that 
the system is indeed in the superfluid phase. More importantly, one 
sees a large enhancement of $|\Delta(z)|$, near $z=0$, compared to 
the uniform system, which shows the formation of bosonic bound 
molecules. A consequence of this lattice induced molecular formation
is that the energetic instability due to pair-breaking excitations 
is suppressed and does not occur at any values of $P$.

For the same parameters of Fig.~\ref{fig:qp_energy2}, the energetic 
instability due to long-wavelength phonons instead occurs at 
$P=0.226\hbar q_{\rm B}=0.452P_{\rm edge}$ and the corresponding 
critical velocity at $1/k_{\rm F}a_s=-1$ is $v_c=0.0662 v_{\rm F}$. 
We also find $v_c=0.0429 v_{\rm F}$ at unitarity and 
$v_c=0.0302 v_{\rm F}$ at $1/k_{\rm F}a_s=1$. This means that, 
in the whole crossover, the critical velocity at $s=5$ is largely reduced 
compared to the red line in Fig.~\ref{fig:vc}(c) for $s=1$ and does
not exhibit a maximum anymore. The behavior of $v_c$ as a function 
of $s$ at unitarity has been already discussed in Ref.~\cite{vc}.

\section{Summary and Outlook\label{summary}}

We have studied the effects of a periodic potential
on the Landau critical velocity of a Fermi superfluid 
in the BCS-BEC crossover. We have considered
a 3D superfluid Fermi gas flowing in a 1D periodic 
potential produced by an optical lattice. 
Using the Bogoliubov-de Gennes equations, we have obtained
a unifying picture both for weak and strong lattices
and in the whole BCS-BEC crossover. We have found that, 
when the recoil energy is comparable to the Fermi energy,
energetic instability due to fermionic pair-breaking excitations
can be less effective as a consequence of the 
periodic structure of the quasiparticle energy spectrum.
When the lattice height is much larger than the Fermi energy,
pair-breaking excitations are prevented because
the lattice potential gives a stronger attraction between paired 
atoms, eventually forming bound bosonic molecules.
We have also found that, when the recoil energy is comparable to 
or larger than the Fermi energy, the critical velocity
due to the long wavelength phonon excitations is drastically reduced
by the lattice in the BCS regime leading to its non-monotonic behavior
along the BCS-BEC crossover.

A further interesting issue regards the possible existence of
roton-like minima in the bosonic dispersion curve.  This excitations
are obtained at low filling fractions and within a tight-binding 
attractive Hubbard model \cite{sofo,kostyrko,ganesh,yunomae}.
The roton-like minima arise from strong charge-density-wave fluctuations.
These fluctuations are expected to be less favored in our 
system, where the gas is uniform in the transverse directions.
However, if such roton-like excitations exist also in the our 
case (3D gas in a 1D lattice), they would lower the critical 
velocity in the BCS regime and for strong lattices. To address this 
issue, one should use, for instance, a quasiparticle 
random phase approximation (QRPA) on top of the
stationary solution of the Bogoliubov-de Gennes equations. 
This is an interesting challenge for future investigations.

Finally, we would like to discuss a similarity between the 
present system and nuclear ``pasta'' phases \cite{rpw,hashimoto,qmd} 
in crusts of neutron stars. The pasta nuclei are those of exotic 
shapes such as rod-like and slab-like structures.
In neutron star crusts, the pasta nuclei
are immersed in background electrons and a gas of dripped neutrons,
which is regarded to be in the superfluid phase.
The setup considered in the present work resembles superfluid neutrons in the
pasta phase with slab-like nuclei, which are in the normal phase and 
provide a 1D periodic potential for superfluid neutrons \cite{note_estimate}.

\acknowledgments
We thank Ki-Seok Kim, Chris Pethick, Robin G. Scott, and Tetsuya Takimoto 
for helpful discussions and comments.
Calculations were performed on the RIKEN Integrated Cluster of
Clusters (RICC) system, WIGLAF at the University of Trento, 
and BEN at ECT*.  
GW acknowledges the Max Planck Society (MPG), 
the Korea Ministry of Education, Science and Technology (MEST), 
Gyeongsangbuk-Do, and Pohang City for the support of 
the Independent Junior Research Group 
at the Asia Pacific Center for Theoretical Physics (APCTP).
This work, as a part of the European Science
Foundation EUROCORES Program EuroQUAM-FerMix, is supported by funds
from the CNR and the EC Sixth Framework Programme. It is also
supported by MiUR.

\bigskip

\appendix

\section{Dynamical instability within hydrodynamic analysis\label{dynamical}}

In this Appendix, we show the results of the critical velocity
$v_c$ for dynamical instability due to long-wavelength excitations.
This instability occurs when a non-zero imaginary part appears
in the excitation energy (\ref{eq:dispersion}). The condition is
\begin{equation}
  \frac{\partial^2 e}{\partial n_0^2}
\frac{\partial^2 e}{\partial P^2}<0 \; .
\label{eq:dyncond}
\end{equation}
Even though the long-wavelength excitations in general do not give 
the lowest critical velocity of the dynamical instability, this 
condition becomes useful at $V_0 \gg E_{\rm F}$, when the $P$-dependence 
of $e$ is almost sinusoidal and the critical velocity 
dependents rather weakly on the wavelength of the excitations.
In this case, the above condition gives an 
onset of the dynamical instability at $P=P_{\rm edge}/2$, which 
coincides with the condition $e(P+\hbar q)-e(P)=e(P)-e(P-\hbar q)$ 
for $V_0 \gg |\Delta|$ \cite{wu}.  This corresponds to the 
energy and momentum conservation for two particles decaying into 
two different Bloch states with $e(P\pm\hbar q)$.

\begin{figure}[!t]
\begin{center}\vspace{0.5cm}
\rotatebox{0}{
\resizebox{8.5cm}{!}
{\includegraphics{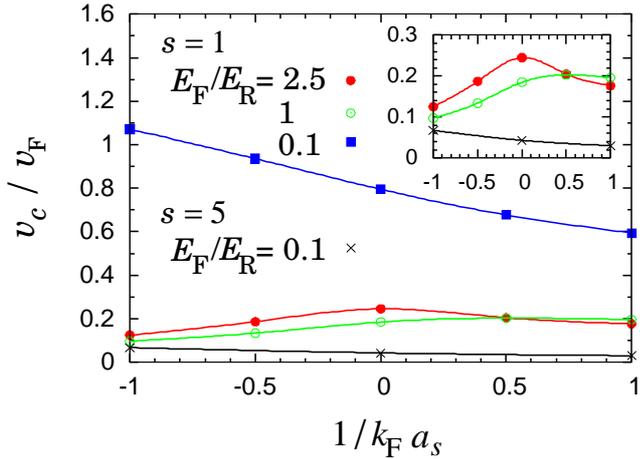}}}
\caption{\label{fig:vcdyn}(Color online)\quad
Critical velocity $v_c$ for dynamical instability
due to long-wavelength excitations in the BCS-BEC crossover.
The solid lines are guides 
to the eye.
The inset shows the magnification of the region of $v_c/v_{\rm F}=0$ -- $ 0.3$.
}
\end{center}
\end{figure}

In Fig.\ \ref{fig:vcdyn}, we show the critical velocity 
for the dynamical instability determined by the 
condition (\ref{eq:dyncond}). The most striking feature is 
the large values of $v_c$ 
in the case of $s=1$ and $E_{\rm F}/E_{\rm R}=0.1$.
As in the case of the energetic instability, this is due to 
the fact that phonons are insensitive to the lattice 
when $s\alt 1$ and $E_{\rm F}/E_{\rm R}\ll 1$, i.e.,
the healing length is much larger than the lattice spacing. 
Note that, even for the same value of $E_{\rm F}/E_{\rm R}=0.1$,
the critical velocity can be rather small provided $s\gg 1$. This
tendency is confirmed by the results for $s=5$ and $E_F/E_R=0.1$ shown 
by crosses in Fig.\ \ref{fig:vcdyn}.

\begin{thebibliography}{99}
\bibitem{note_assumption} Here, we ignore the possible  
existence of roton-like minima in the collective excitation spectrum
obtained in the BCS regime for 1D (2D or 3D) optical lattice 
in a 1D (2D or 3D) system using a tight-binding attractive Hubbard model 
\cite{sofo,kostyrko,ganesh,yunomae}.  See also Sec.\ \ref{summary}.
\bibitem{note_partial} Note that the partial derivatives 
in Eq.\ (\ref{eq:vcrit-hydro}) should be done by keeping 
$a_s$ and $d$ constant.
\bibitem{note_sound} Note that the sound velocity at $P=0$ 
basically decreases
with increasing $1/k_{\rm F}a_s$ for all values of 
$E_{\rm F}/E_{\rm R}$. However, the velocity at which the energetic 
instability occurs due to the long-wavelength phonon excitations 
does not show a monotonic behavior with increasing 
$1/k_{\rm F}a_s$ for $E_{\rm F}/E_{\rm R}=2.5$ and $1$, unlike 
$E_{\rm F}/E_{\rm R}=0.1$.
\bibitem{note_estimate} In the case of the nuclear pasta phase 
in neutron stars, an estimate of the critical velocity can be 
obtained as follows.  For the pasta phase with slab-like nuclei,
the average neutron density $n_0\simeq 0.08$~fm$^{-3}$ and 
the lattice constant $d\simeq 15$ -- 20 fm and thus 
$E_{\rm F}/E_{\rm R}=(k_{\rm F}/q_{\rm B})^2 \sim 40$ -- 70.
The strength $V_0$ of the optical lattice corresponds to the
depth of the bottom of the conduction band of neutrons 
measured from the average potential energy outside nuclei \cite{pethick}.
Since the neutron Fermi energy $E_{\rm F,\, in}$ for the density of
neutrons inside nuclei is $E_{\rm F,\, in}\sim 35$ -- 40 MeV and the
neutron chemical potential $\mu_{\rm in}$ inside nuclei is 
$\mu_{\rm in} \sim 10$ -- 15 MeV, 
we estimate $V_0 \sim |\mu_{\rm in}-E_{\rm F,\, in}| \sim 25$ MeV.  
Thus, we obtain $s\sim 25$ -- 50.
Since the absolute value of the neutron scattering length $a_s\simeq -18.5$~fm
is much larger than the radius $r_s$ of the average volume per neutron,
$r_s\simeq 1.5$~fm, superfluid neutrons in neutron star crusts 
are roughly approximated as a unitary Fermi gas 
\cite{note:re,bertsch,baker} (see also Refs.~\cite{gezerlis,abe,pethick}). 
Here, because $E_{\rm F}/E_{\rm R}\gg 1$, we can safely employ the 
hydrodynamic theory with the local density approximation given 
in Ref.~\cite{vc} to estimate the critical velocity.  
Using Eq.~(3) of Ref.~\cite{vc} together with $\beta=-0.58$ \cite{beta} and 
the above parameters, we estimate $v_c\sim 10^{-3}$ -- $10^{-2} v_{\rm F}$
with $v_{\rm F}\simeq 0.3 c$, where $c$ is the speed of light.
\bibitem{note:re} Here, just to give an order of estimate, 
we shall ignore a quantitatively significant correction 
by the effective range $r_e\simeq 2.7$~fm \cite{schwenk}, 
which is comparable to the interparticle separation between neutrons.
\end{thebibliography}
\end{document}